\documentstyle[multicol,graphics,aps]{revtex}       %draft,overcite,

\title{Effects of Electron-Electron Scattering on
Electron-Beam Propagation in a Two-Dimensional Electron-Gas}

\author{H.\ Predel, H.\ Buhmann, and L.W.\ Molenkamp}
\address{Physikalisches Institut der Universit\"at W\"urzburg,
D-97074 W\"urzburg, Germany}
\author{R.N.\ Gurzhi, A.N.\ Kalinenko, A.I.\ Kopeliovich, and A.V.\
Yanovsky}
\address{B.\ Verkin Institute for Low Temperatures Physics \&
Engineering,\\
 Nat.Acad of Sciences of Ukraine, Lenin Ave. 47,
31064, Kharkov, Ukraine}

\begin{document}

\draft \preprint{H.\ Predel}

%\date{\today}

\maketitle

\begin{abstract}
We have studied experimentally and theoretically the influence of
electron-electron collisions on the propagation of electron beams
in a two-dimensional electron gas for excess injection energies
ranging from zero up to the Fermi energy. We find that the
detector signal consists of {\em quasiballistic} electrons, which
either have not undergone any electron-electron collisions or have
only been scattered at small angles. Theoretically, the
small-angle scattering exhibits distinct features that can be
traced back to the reduced dimensionality of the electron system.
A number of nonlinear effects, also related to the two-dimensional
character of the system, are discussed. In the simplest situation,
the heating of the electron gas by the high-energy part of the
beam leads to a weakening of the signal of quasiballistic
electrons and to the appearance of thermovoltage. This results in
a nonmonotonic dependence of the detector signal on the intensity
of the injected beam, as observed experimentally.

\end{abstract}

\pacs{72.10.-d;72.20.Dp}

%%%%%%%%%%%%%%%%%%%%%%%%%%
\begin{multicols}{2}
%%%%%%%%%%%%%%%%%%%%%%%%%%%

\section{Introduction}

The propagation of electron beams in the two-dimensional
electron-gas (2DEG) of GaAs-(Al,Ga)As heterostructures was studied
in a number of publications \cite {Sprl,Pprl,Msst,Juelich,Appl},
and has proven to be a very sensitive tool for studying electron
scattering-phenomena. In the first two Refs.\ \cite {Sprl,Pprl},
the emphasis was on the effects of {\em electron-phonon}
scattering, where the beam was injected across tunnel barriers.
These effects occur at relatively large excess energies of the
electron beam, typically of the order of the optical phonon
energy, some 30 meV. In later works \cite{Msst,Juelich}, the
effects of {\em electron-electron} scattering phenomena (occurring
at much lower energies, typically below 10 meV) were analyzed,
using opposite quantum point-contacts as injector and detector for
the electron beam.

In our paper, Ref.~\cite {Msst}, we paid much attention to thermal
beams in which the characteristic energy of beam electrons
$\varepsilon$, counted from Fermi level, is of the order of the
sample temperature $T_0$. It was shown that electron-electron
collisions played a main role in damping such beams. The overall
behaviour of the signal attenuation could be reasonably understood
using the formula of Giuliani and Quinn \cite{Quinn} for the
electron-electron scattering rate in a 2DEG, implying that the
result of a single electron-electron collision is sufficient for
an unequilibrium electron to escape from detection (relaxation
time approximation). This conclusion was subsequently confirmed by
other groups \cite{Juelich,Appl}.

In this work, we return to our studies of electron-electron
scattering in a 2DEG system, equipped with a much more detailed
framework of understanding of the dynamical scattering phenomena
\cite{Gprl,Gprb,Glt1,Gssc,Glt2}, which has first been used to
explain the hydrodynamic electron flow-phenomena we observed a few
years ago \cite{deJong}. These newly developed theories enable a
much more refined analysis of the experimental data. Specifically
targeting the theoretical predictions of Refs.\
\cite{Gprl,Gprb,Glt1,Gssc,Glt2} we have performed a new series of
electron-beam experiments for different samples at various
temperatures and for a wide range of injection energies. In our
experiments we can identify specific two-dimensional effects, as
well as novel nonlinearities due to 2DEG-heating. Our results cast
doubts on the interpretations in Refs.\ \cite{Juelich} and
\cite{Appl}.

In the course of this paper, we will first present the
experimental results and their qualitative explanation (Sec.\
\ref{sec_exp}). Next we develop a theoretical approach for the
electron-beam propagation in small systems, i.e.\ where the
probability of secondary collision is negligible (Sec.\
\ref{sec_onecoll}), and for the opposite case, the multi-collision
limit (the propagation of a beam over long distances becomes
possible due to specific two-dimensional effects) (Sec.\
\ref{sec_multicoll}). In Sec.\ \ref{sec_heat} we consider
non-linear phenomena which can play an essential role for the
interpretation of an electron-beam signal and we analyse the
experimental data in the framework of the here developed theory in
Sec.\ \ref{sec_discuss}. Throughout this paper we will use
'energy-units' for temperature and potential differences, i.e. the
Boltzmann constant $k_B$ and the electron charge $e$ are equal to
one.

\section{Experiment}\label{sec_exp}

The experiments were performed on gate-defined nanostructures in
conventional modulation doped GaAs-(Al,Ga)As heterojunctions.
Typical values for the carrier density and mobility are $n_e=2.45
\times 10^{11}$ cm$^{-2}$ and
\begin{figure}[t]
\begin{center}
\resizebox{7cm}{8.8cm}{\includegraphics{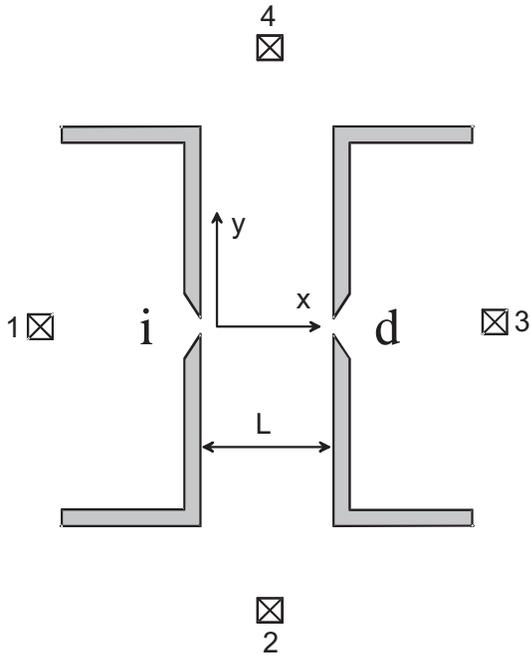}}
\begin{minipage}{8cm}
\caption{\label{sample} Schematic topview of the device layout.
Hatched areas are Schottky gates, defining the device geometry,
crossed squares denote Ohmic contacts, $i$ and $d$ symbolize the
injector and detector quantum point-contact, respectively.}
\end{minipage}
\end{center}
\end{figure}
$\mu \approx 1\times 10^6$
cm$^2$(Vs)$^{-1}$, corresponding to an impurity mean-free-path of
$l_{\rm imp} \approx$ 20 $\mu$m. A schematic topview of the sample
gate-structure is given in Fig.\ \ref{sample}. Schottky gates
(grey areas) form two opposite quantum point-contacts, $i$ and
$d$, at lithographical distances of $L = $ 0.6, 2, 3.4, and 4
$\mu$m for different samples.

In the experiments, the electron beam was injected through the
injector quantum point-contact $i$ by applying a dc voltage
$V_i=V_{12}$ and detected as the non-local voltage $V_d=V_{34}$
across the detector point-contact $d$. The numbers 1, 2, 3, and 4
denote the Ohmic contacts to the 2DEG of the sample (Fig.\
\ref{sample}, crossed squares). We stress that the use of all-dc
techniques is very important for a proper interpretation of the
observed signals. Differential resistance measurements with
lock-in techniques will not elucidate the role of the
thermovoltage background to the signal in full. Both injector and
detector point-contacts were adjusted at the $n=1$ plateau i.e.,
both contain one transverse mode, and thus remain in the metallic
regime, $G_{\rm QPC} = n 2e^2/h$. (In other words, they do not act
as tunnel barriers, as was the case in Refs.\ \cite{Sprl,Pprl}.)
Thus, electrons of all possible energies, $0 < \varepsilon < V_i$,
are present in the injected beam.

\begin{figure}[t]
\begin{center}
\resizebox{7cm}{4.7cm}{\includegraphics{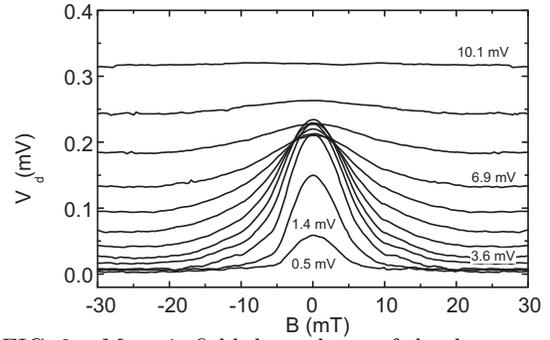}}
\begin{minipage}{8cm}
\caption{\label{beam} Magetic-field dependence of the detector
signal $V_d$ for different injection voltages: 0.5, 1.4, 2.2, 2.9,
3.6, 4.4, 5.5, 6.0, 6.9, 7.8, 8.8 , and 10.1 mV (bottom to top).
Note that no offset is added to the experimental data.}
\end{minipage}
\end{center}
\end{figure}

In the presence of a magnetic field perpendicular to the 2DEG
plane, the electron beam is deflected and the detector signal,
$V_d(B)$ yields the beam profile (see Fig.\ \ref{beam}). For the
present a point-contact adjustment at $n=1$ and a
injector-detector distance $L=3.4$ $\mu$m we obtain the
characteristic opening angle of injector and detector which
amounts to $\phi \sim 18^{\circ}$ (cf.\ \cite{Mprb}).

From Fig.\ \ref{beam}, showing the $V_d(B)$-dependence for
different injection energies ($0.5\le V_i \le 10$ mV) at 1.6 K,
one can see that the detector signal first increases with
increasing injector voltage. Then for $V_i > 3$ mV a strong
increase of a isotropic background signal is observed while at the
same time the beam profile broadens. For injection energies larger
10 meV a beam signal can hardly be resolved, while the background
increases continuously.

\begin{figure}[t]
\begin{center}
\resizebox{7cm}{12.25cm}{\includegraphics{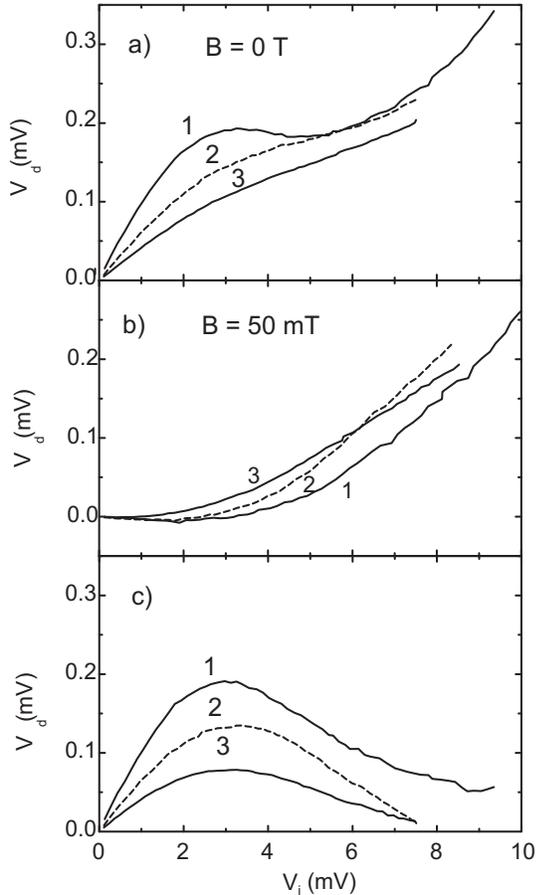}}
\begin{minipage}{8cm}
\caption{\label{exp} Experimental dependence of the detector
signal $V_{d}$ on the injector voltage $V_{i}$ for  $L = 3.4$
$\mu$m at three different sample temperatures $T_0 = 1.6$ K (trace
1), 8.0 K (trace 2), and 11 K (trace 3). (a) detector signal at
zero magnetic field - electron beam is directed straight from
injector to detector, (b) $B = 50$ mT - the electron beam is
deflected and does not reach the detector directly [isotropic
background signal (thermovoltage)]. (c) Contribution of
narrow-directed (quasiballistic) movement of electrons to the
detector signal resulting from subtracting the appropriate data
sets of Fig. 2 b) from Fig. 2 a) [$T_0 = 1.6$ K (trace 1), 8.0 K
(trace 2), and 11 K (trace 3)].}
\end{minipage}
\end{center}
\end{figure}

To investigate the effects of electron-electron scattering events
on the beam propagation we are interested in the dependence of
detector-signal on the injection energy at $B=0$. Fig.\ \ref{exp}
a) presents the experimental results for the sample $L=3.4$ $\mu$m
at three different sample temperatures, $T_0$ = 1.6, 8 and 11 K.
Additional measurements (not shown here) were made at different
sample (lattice) temperatures, $T_0$ = 2.2, 3.4, 5, 15, and 17 K
and for different injector-detector distances. It can be seen from
Fig.\ \ref{exp} a) that for low injection energies the detector
signal increases linearly with $V_i$. For $V_i > 3$ meV only for
the lowest temperature (curve 1) a saturation and even a small
decrease is observed. For high injection energies the $V_d(V_i)$
dependence increases for all temperatures.

As we have seen from Fig.\ \ref{beam}, for $V_i > 3$ meV an
increasing isotropic background signal is detected, which is not
directly related to the ballistic electron-beam propagation. In
order to extract the ballistic part of the detector signal we
measured the isotropic background signal separately by repeating
the experiment for high magnetic fields ($B=50$ mT) to ensure that
the electron beam is totally deflected and ballistic beam
electrons do not contribute to the detector signal [Fig.\
\ref{exp} b)]. Subtracting this background signal from the data
measured for $B=0$ T we obtain the pure electron-beam contribution
to the detector signal [Fig.\ \ref{exp} c)]. Now the result is
similar for all temperatures: We observe first a linear increase
of $V_d$ with increasing $V_i$ and then a saturation followed by a
decrease for high injection energies, while with increasing sample
temperature the maximum electron-beam signal decreases [Fig.\
\ref{exp} c)].

\begin{figure}[t]
\begin{center}
\resizebox{7cm}{5cm}{\includegraphics{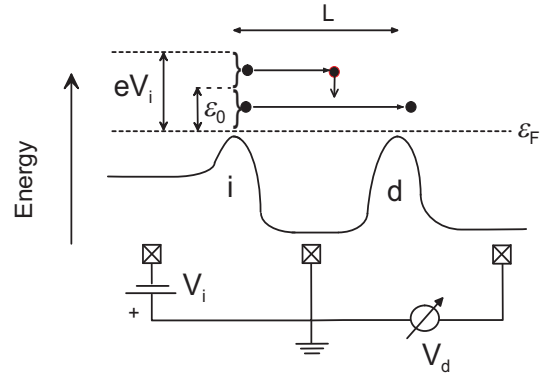}}
\begin{minipage}{8cm}
\caption{\label{pot}Schematic view of the characteristic energies
of the electron-beam experiments. Potential barriers are shown at
the center of the quantum point-contacts. Typical energies are
indicated: $\varepsilon_F$ (Fermi energy), $eV_i$ (injection
energy), and $\varepsilon_0$ [maximum energy for electrons
reaching the detector ballistically ($l(\varepsilon_0)=L$)]. Note,
the effective charging of the area behind the detector ($eV_d$) is
at least 30 times smaller than the injection energy and therefore
negligible on the scale of this diagram.}
\end{minipage}
\end{center}
\end{figure}

These experimental results can be understood from the following
qualitative considerations. Let us assume, for simplicity, that
the lattice temperature (the primary temperature of the system) is
equal to zero. Then, for a nonequilibrium electron with excess
energy $\varepsilon $ above the Fermi energy $\varepsilon _{F}$,
the mean-free-path for collisions with equilibrium electrons
decreases with increasing $\varepsilon$, roughly speaking as
$\l_{ee}( \varepsilon ) \sim \varepsilon^{-2} \ln
\varepsilon^{-1}$, $\varepsilon \ll \varepsilon _{F}$
\cite{Chaplik}. Therefore, at sufficiently small $V_{i}$ all
injected electrons will reach the detector, whose readout then is
proportional to the number of injected electrons, $V_{d} \propto
V_{i}$, schematically shown in Fig.\ \ref{pot}. (Electrons of all
energies $0\leq \varepsilon \leq V_{i}$ are present in the beam,
with equal weight). This linear increase of $V_{d}$ with $V_{i}$
saturates for energies $\varepsilon \ge \varepsilon_0$ when the
electron-
electron scattering mean free path length ($l_{ee}$) becomes
comparable to $L$, the distance between injector and detector:
$V_{i} = \varepsilon _{0}$ for $\l _{ee}\left( \varepsilon
_{0}\right) =L$. Electrons with larger energies, $\varepsilon _{0}
< \varepsilon <V_{i}$, will scatter and do not reach the detector.
Thus, the signal $V_{d}$ is determined by an fraction of electrons
which is completely saturated at $V_{i}\sim \varepsilon _{0}$ and
should not change on a further increase of $V_{i}$. However, as is
evident from Fig.\ \ref{exp} a), the signal, upon reaching a
maximum, starts to decrease slightly. The only possible mechanism
leading to such behaviour is heating of the 2DEG in between
injector and detector point-contact by the electron beam. The
heated 2DEG then leads to damping of the electron beam due to
enhanced electron-electron scattering. At still higher $V_{i}$,
$V_{d}$ shows again an increase [Fig.\ \ref{exp} a)]. This is due
to the additional build-up of a thermovoltage across the detector
point-contact [Fig.\ \ref{beam} and \ref{exp} b)], which is driven
by the temperature difference between the heated 2DEG in between
injector and detector and the still cold 2DEG behind the detector
\cite{thermo}.

As we will see below, the qualitative picture given above is fully
confirmed by the theory described in this work. We will
demonstrate that under our experimental conditions it is possible
to separate the electrons of the beam into two groups, i.e.
"quasiballistic" ($\varepsilon <\varepsilon _{0}$) and "heating
electrons" ($\varepsilon >\varepsilon _{0}$), which greatly
simplifies the interpretation of the experimental results.

\section{One-Collision Approximation}\label{sec_onecoll}

The detected signal $V_{d}$ is determined by the distribution
function of nonequilibrium electrons $f$ in the vicinity of
detector point-contact. For now, we neglect nonlinearities due to
heating of the 2DEG in between injector and detector, which is a
valid approximation for sufficiently low excess energies of the
injected electrons. The linearized Boltzmann equation, describing
the behaviour of the distribution function $f$, then has the form

\begin{equation}
v_{x}\frac{\partial f}{\partial x}+v_{y}\frac{\partial f}{\partial
y}=\widehat{J}f, \hspace{5mm}
f\left( x=0,y,{\bf p}\right) =f_{0}\left(y,{\bf p}\right)  %\tag{Eq1}
\end{equation}

Here, $f_{0}\left( y,{\bf p}\right) $ is the beam profile at the
exit from the injector, and the axis $x$ is directed from injector
to detector (cf.\ Fig.\ \ref{sample}). $\widehat{J}f$ is a
linearized integral describing the electron-electron collisions.
It is convenient to write it as

\begin{equation}
\widehat{J}f=-\nu f+\int d{\bf p}^{\prime }\nu _{{\bf pp}^{\prime }}
f_{{\bf p}^{\prime }} \ , \hspace{5mm}
\nu =\int d {\bf p}^{\prime }\nu _{ {\bf p}^{\prime }{\bf p}} \ . %\tag{Eq2}
\end{equation}

Here, the collision-integral kernel $\nu _{{\bf p}^{\prime }{\bf
p}}$ determines the probability of the appearance of a
nonequilibrium electron ($\nu_{{\bf p'p}} > 0$) or hole
($\nu_{{\bf p'p}} < 0$) in state ${\bf p}^{\prime }$ after the
nonequilibrium electron has disappeared from state ${\bf p}$
(i.e., has been scattered into another state). The kernel has a
complex structure and in the general case can not be presented in
elementary functions. We have

\begin{equation}
\nu _{{\bf p}^{\prime }{\bf p}}=\frac{1}{n\left( \varepsilon \right)}
\int d{\bf p}_{1}d{\bf p}_{2}(2\ \Psi _{{\bf p}^{\prime }{\bf p}_{1}
{\bf pp}_{2}} - \Psi _{{\bf p}^{\prime }{\bf pp}_{1}{\bf p}_{2}})
\hspace{5mm} ,%\tag{Eq3}
\end{equation}
where

\begin{eqnarray}
\Psi _{{\bf p'p}{\bf p}_{1}{\bf p}_{2}}= W_{{\bf p'p}{\bf p}_1
{\bf p}_2} (1-n\left( \varepsilon' \right) ) n\left(
\varepsilon _{1}\right) n\left(\varepsilon_{2}\right) \times
\nonumber\\
\times \delta \left( {\bf p'}+{\bf p}-{\bf p}_{1}-{\bf p}_{2}
\right) \delta \left( \varepsilon' +\varepsilon -
\varepsilon_{1}-\varepsilon _{2}\right). %\tag{Eq4}
\end{eqnarray}
Here $W_{{\bf p'p}{\bf p}_{1}{\bf p}_{2}}$ is proportional to the
square of the matrix element of the electron-electron interaction,
and $n\left(\varepsilon \right)$ is the equilibrium Fermi
distribution function.

We now introduce the angular scattering distribution function:

\begin{equation}
g\left( \varphi \right) =\nu ^{-1}m\int d\varepsilon ^{\prime }
\nu_{{\bf p}^{\prime }{\bf p}} \, , %\tag{Eq5}
\end{equation}

where $\varphi$ is the scattering angle. For simplicity, we assume
parabolic bands ($\varepsilon =p^{2}/2m$), which is a good
approximation for the conduction band in GaAs-(Al,Ga)As
heterostructures. At sufficiently small $\varepsilon $ and $T$,
the form of $g\left( \varphi \right)$ is determined mainly by the
phase-space restrains imposed by the two-dimensional character of
the 2DEG \cite {Gprl,Gprb,Glt1,Gssc}. Roughly speaking, $g\left(
\varphi \right)$ consists of a narrow bunch of electrons flying
forward in an angle range of the order of $\pm \left( \varepsilon
+T\right)^{1/2}\varepsilon _{F}^{-1/2}$ and a bunch of holes, of
approximately the same width, flying backward (see Ref.\
\cite{Glt2}). Therefore, the electron-electron scattering is
effectively a small-angular process \cite{Gprl,Gprb,Glt1}.

For the general case, Eq. (1) can not be solved. However, under
conditions where the probability of collisions is small, i.e.
$\l_{ee}=v( \varepsilon ) \nu^{-1} ( \varepsilon ) \gg L$, or
$\varepsilon \ll \varepsilon _{0}$, we can use perturbation theory
for the collision integral. In the first order or one-collision
approximation we then have

\begin{eqnarray}
&f\left( x,y,{\bf p}\right)= \left( 1-\frac{x\nu }{v_{x}}\right)
f_{0}\left( y-\frac{v_{y}}{v_{x}}x,{\bf p}\right) +\nonumber\\
  &  +\frac{1}{v_{x}} \int\limits_{0}^{x}dx^{\prime } \int d{\bf
p}^{\prime }\nu _{{\bf p p'}} \ f_{0}\left[
y-\frac{v_{y}}{v_{x}}x+\left( \frac{v_{y}}{
v_{x}}-\frac{v_{y}^{\prime }}{v_{x}^{\prime }}\right) x^{\prime
},{\bf p'} \right] \equiv \nonumber\\
    &\equiv \left( 1-\frac{x\nu
}{v_{x}}\right) \ f_{0}+\widehat{Q}f_{0}.
\end{eqnarray}

The first term on the r.h.s of Eq.\ (6) describes the number of
nonscattered particles reaching into the vicinity of a point
$\left( x,y\right) $. The second (integral) term
$\widehat{Q}f_{0}$ describes particles that reach the same spatial
region, after having been scattered once.

Note that for high-energy beams ($\varepsilon \gg T$), the
probability of undergoing a second collision is approximately one
order of magnitude lower than that of the first collision \cite
{Gssc}. This is connected with the fact that after collision with
equilibrium (Fermi sea) electrons, the excess energy of a
nonequilibrium electron ($\varepsilon $) must be redistributed
between three partners i.e., $\bar{\varepsilon }\approx
\varepsilon /3$, $\l _{ee} (\bar{\varepsilon}) \approx 3^{2}\l
_{ee} ( \varepsilon )$, $T\ll \varepsilon \ll \varepsilon _{F}$,
where $\bar{\varepsilon }$ is the characteristic energy of the
scattered electrons. Therefore, the one-collision approximation is
valid for a relatively wide range of energies as long as $\l _{ee}
( \bar{\varepsilon} ) \ge L$. On observing this and the fact that
$\widehat{Q}\ \nu \ f_{0}\sim \nu \left( \varepsilon \right)
\widehat{Q}\ f_{0}\gg \nu \ \widehat{Q}\ f_{0}\sim \nu \left(
\bar{\varepsilon } \right) \ \widehat{Q}\ f_{0}$, it is
straightforward to build a new "modified" one-collision
approximation. After partial summation of the terms of the
iteration series on the parameter $x/l_{ee}(\varepsilon)$ of Eq.
(1) one obtains the following expression in zero-eth order
approximation for the parameter
$x/l_{ee}\left(\bar{\varepsilon}\right) $:

\begin{eqnarray}
& f \approx e^{ -\frac{\nu \ x}{v_{x}}}\ f_{0}\left( y-\frac{%
v_{y}}{v_{x}}x,\ {\bf p}\right) +\nonumber\\
  & + \frac{1}{v_{x}}\int\limits_{0}^{x}dx^{\prime }
    \int d{\bf p}^{\prime }\nu _{{{\bf pp}}^{\prime}} e^{ -\frac{\nu(
    \varepsilon ^{\prime }) \ x^{\prime}} {v_{x}^{\prime}}}\nonumber\\
 &\times f_{0}\left[ y-\frac{v_{y}}{v_{x}}x+\left( \frac{v_{y}}{v_{x}}-%
\frac{v_{y}^{\prime }}{v_{x}^{\prime }}\right) x^{\prime },\ {\bf p}
^{\prime }\right] \, .%\tag{Eq7}
\end{eqnarray}

This formula is valid when $\varepsilon < 3\ \varepsilon_{0}$. The
first term on the r.h.s. corresponds to the usual relaxation-time
approximation, $\widehat{J}\ f=-\nu\ f$. Note that the modified
one-collision approximation Eq.\ (7) is based on an exact
consideration of the first collision and not on perturbation
theory. It does not take into account any further collisions.

The experimentally measured voltage drop, $V_d$, is determined by
the current passing the detector point-contact and can be
calculated from

\begin{equation}
V_{d}=e \int d\varepsilon \int d\varphi \  \rho \left( \varphi \right)
v_{x}\ f\left( x=L,y=0,{\bf p}\right) \, .
%\tag{Eq8}
\end{equation}

Here $\rho \left( \varphi \right) $ is the function characterizing
the angular acceptance of the detector point-contact, which is
positioned at $(L,0)$. For simplicity, we use in our numerical
calculations Heaviside step-functions to represent the angular
characteristics of injector and detector point-contacts:

\begin{figure}[t]
\begin{center}
\resizebox{7cm}{5cm}{\includegraphics{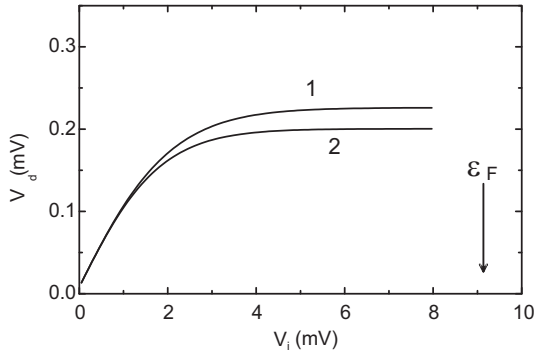}}
\begin{minipage}{8cm}
\caption{\label{onecoll} Calculated $V_{d} (V_{i})$ dependences
without taking into account electron heating of the 2DEG in
between injector and collector. Curve 1: Modified one-collisional
approximation Eqs.(7),(8); curve 2: relaxation-time approximation;
with $\phi =18^{\circ}$, $L = 3.4$ $\mu$m, $\varepsilon_{0} =0.2$
meV, and $\varepsilon_{F} = 9$ meV.}
\end{minipage}
\end{center}
\end{figure}

\begin{eqnarray}
\rho \left( \varphi \right)  &\propto &\theta \left( \phi /2-\left|
\varphi \right| \right)\nonumber
%\tag{Eq9}
\\
\vphantom{a}\\
f_{0}\left( y,{\bf p}\right)  &\propto &\theta \left( \phi /2-\left|
\varphi \right|
\right) \theta \left( V_{i}-\varepsilon \right) \, . \nonumber
\end{eqnarray}

In this model, the behaviour of $V_{d}\left( V_{i}\right)$ is
determined by two parameters, i.e. the angular injection (and
acceptance) range of the point-contact $\phi$ and the distance
between injector and detector $L$. For more realistic models of
the angular response of quantum point-contacts we refer to Ref.\
\cite{Mprb}.

The dependence $V_d(V_i)$ is calculated using Eqs.\ (7-9),
including the expressions for the kernel $ \nu_{\bf pp'}$ obtained
in Ref.\ \cite{Glt2}, setting $\phi =18^{\circ}$, and $L$ = 3.4
$\mu$m, i.e. close to the experimental conditions. The result is
shown in Fig.\ 5 for the full expression of the modified
one-collision approximation (MOCA) (curve 1) and the
relaxation-time approximation (RTA) (curve 2). We clearly observe
that the curves saturate with increasing $V_i$. Saturation occurs
at a higher injection voltage, $V_i$, and a higher signal level,
$V_d$, for the MOCA as compared with the RTA. The difference
between these curves (about 15 \%) characterizes the role of
two-dimensional effects for the given parameters. This difference
is due to the integral term in Eq.\ (7), which can be omitted when
2D effects are negligible.

In the next section we show that the role of the
two-dimensionality is much larger when $\phi >\left(\varepsilon
_{0}/\varepsilon _{F}\right) ^{1/2}$. In this limit, a saturation
of the curve at $V_{i}<3\ \varepsilon _{0}$ does not take place at
all.

\section{Multi-collision regimes}\label{sec_multicoll}

In the limit where the electrons undergo a number of collisions on
their way from injector to detector, it is impossible to obtain a
completely analytical solution of the spatially-inhomogeneous
problem of beam propagation. Instead, we will discuss below a
simple qualitative theory that adequately describes this
multi-collision regime.

\begin{figure}[t]
\begin{center}
\resizebox{7cm}{9.2cm}{\includegraphics{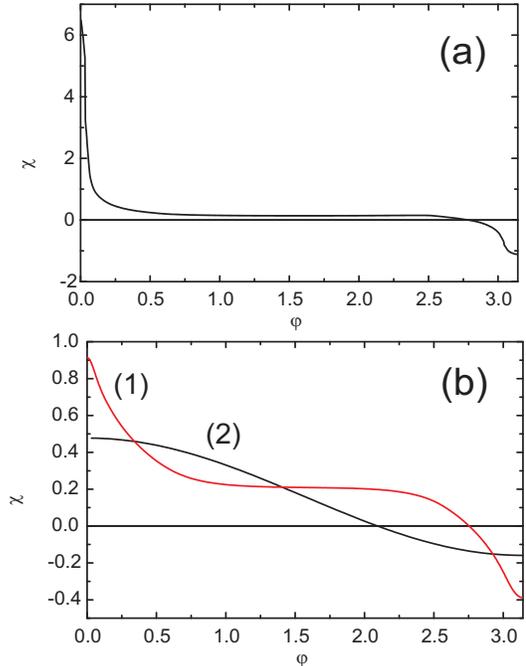}}
\begin{minipage}{8cm}
\caption{\label{multicoll} The temporal evolution of a thermalized
spatially-uniform distribution $\chi (t, \varphi)$
[$T/\varepsilon_{F} = 0.1$ and $\chi |_{t=0} \propto \delta
(\varphi)$]. (a) $t=\tau_{ee}$; (b) $t=10 \ \tau_{ee}$ (curve 1).
Curve 2 for comparison the 3D case after a few electron-electron
collisions (drift-like distribution).}
\end{minipage}
\end{center}
\end{figure}

To obtain realistic numerical values of the angular relaxation
rate we first consider the momentum relaxation in time for a
spatially homogeneous distribution. For simplicity, we take the
thermalized distribution $f=(-\partial n/\partial \varepsilon)
\chi(\varphi,t)$, i.e. equilibrium is established in energy but
not in momentum. In this case the kernel of the collision integral
contains only differences in the angular variables, and the
solution of the Boltzmann equation reduces to the calculation of a
one-dimensional Fourier transform. Here, we use the numerical
results of the angular distribution function $g(\varphi)$ that
were obtained in Ref.\ \cite{Glt2}. (In the case of
non-thermalized distributions the angular and energy variables are
not separable and therefore the solution of the Boltzmann equation
becomes a much more difficult problem.) Fig.\ \ref{multicoll}
shows the results of a calculation for thermalized conditions at
$T_0 = 0.1\ \varepsilon_F$ and for different times $t$ after beam
injection. From this figure follows that the beam remains narrow
up to times of the order of $10\ \tau_{ee}$, whereas in the
three-dimensional case a smooth drift-like distribution is already
established after a delay of the order of one collision-time. From
now on, we will use the convention $0.1\ \varepsilon_F \equiv
\varepsilon^{*}$ to denote the characteristic energy of a beam,
below which the specific features of two-dimensional relaxation
essentially manifest themselves.

Depending on the relative magnitude of $\varepsilon_0$ and the
temperature $T$ of the 2DEG, different multi-collision regimes are
possible:

1. Let us start with the case of low temperatures, $T \ll
\varepsilon_0$. We assume that $L$ is so large that $\varepsilon_0
< \varepsilon^{*}$. In this case, the particles that undergo
multiple collisions but still contribute to the electron beam
signal, are those whose mean-free-path is considerably less than
$L$ and whose scattering is small-angular:

\begin{equation}
\varepsilon_0 < \varepsilon < \varepsilon^{*}\, .
\end{equation}

Note that after a few collisions the energy of such particles
drops very rapidly to values close to $\varepsilon_0$, upon which
the particles will reach the detector without further collision.
In contrast, the opening angle broadening $\alpha$ of the electron
beam is determined by the first collision $\alpha \sim
\sqrt{\varepsilon/\varepsilon_F} \ll 1$. It is then
straightforward to evaluate the contribution to the detector
signal of the group of electrons with energies in the range
$(\varepsilon_0,\varepsilon)$:

\begin{equation}
V_d\sim (\varepsilon-\varepsilon_0) \frac{\lambda_F}{r_{\perp}}
\frac{\phi}{\alpha} \sim \varepsilon_{F} \frac{\varepsilon-
\varepsilon_{0}}{\varepsilon}
\frac{\lambda_F}{L}\phi \; .
\end{equation}

The transverse beam broadening is $r_{\perp}\simeq L\alpha$. The
detector width is chosen to be of the order of the electron Fermi
wave-length $\lambda_F$ (corresponding to the occupation of one
mode in the quantum point-contact). We have assumed an angle of
acceptance $\phi \ll \alpha$ for the detector point-contact
\cite{Mprb}; in the other case, if $\phi \ge \alpha$, the
multiplier $\phi/\alpha$ can be omitted for the above expression.

The order of magnitude of the contribution of ballistic electrons
to the detector signal can be estimated as $V_d\sim \varepsilon_0
\lambda_F (L\phi)^{-1}$, where we assume identical characteristics
for injector and detector quantum point-contacts. Therefore, the
condition for a predominance of the group of non-ballistic
electrons to the detector signal takes the form

\begin{equation}
\phi > \sqrt{\frac{\varepsilon_0}{\varepsilon_F}}\; .
\end{equation}

This inequality is satisfied more easily for samples with larger
$L$ (i.e.\ smaller $\varepsilon_0$) or larger acceptance angles
$\phi$. Under our experimental conditions the l.h.s.\ and r.h.s.\
in Eq.\ (12) coincide by an order of magnitude, and it turns out
that the 2D effects lead to corrections of the order of unity. In
case $\phi \gg \sqrt{{\varepsilon_0}/{\varepsilon_F}}$ it should
be possible to observe the long-distance beam propagation as
predicted in Ref.\ \cite{Gprb}. In other words, one can detect an
electron beam over a distance exceeding substantially the
electron-electron mean-free-path $l_{ee}$ as a result of
one-dimensional electron-hole diffusion.

2. For $\varepsilon_0 \ll T$, ballistic electrons are practically
absent. Roughly speaking, the number of quasi-particles reaching
the detector without any collisions is exponentially small and
proportional to $\exp[-L/l_{ee}(T)]=\exp(-T^2\varepsilon_0^{-2})$.
High-energy electrons with energies $\varepsilon_F >
\varepsilon>T$ loose their excess energy very quickly, after a few
collisions, and cool down to energies of the order of the lattice
temperature $T$. Simultaneously, the beam acquires an angular
broadening of the order of $\sqrt{\varepsilon/\varepsilon_F} <1$.
After this initial relaxation, provided $T< \varepsilon^{*}$, we
still have a narrow distribution of electrons (and holes with
opposite momenta) whose movement is a one-dimensional diffusion in
coordinate space -- an effect which is genuinely caused by the
two-dimensionality of the electron system. The angular broadening
in time of this specific group can be expressed as $\alpha \sim
\sqrt{T/\varepsilon_F}\ [t/\tau_{ee}(T)]^{1/4}$ (cf.\
\cite{Gprl,Gssc}). The contribution of this narrowly-directed
group of electrons to the detected signal can be evaluated using
Eq.\ (11). Taking into account, that, assuming one-dimensional
diffusion, the time an electron needs to reach from injector to
detector is of the order of $v_F^{-1}L^2 l_{ee}^{-1}(T)$ we obtain

\begin{eqnarray}
\alpha \sim \sqrt{\varepsilon/\varepsilon_F}+
\sqrt{T/\varepsilon_F}\left[L/l_{ee}(T)\right]^{1/2},\nonumber\\
\vphantom{a}  \\
r_{\perp}\sim L \left\{ \sqrt{\varepsilon/\varepsilon_F}+
\sqrt{T/\varepsilon_F}\left[L/l_{ee}(T)\right]^{3/2}
\right\}.\nonumber
\end{eqnarray}

As one can see from these expressions, the result of Eq.\ (11)
obtained above is retrieved for $\varepsilon
>T^3\varepsilon_0^{-2}\equiv T^{*}$. At the same time, the
contribution to the detector signal of electrons with energies
$T_0 < \varepsilon < T^{*}$ is given by

\begin{equation}
V_d \sim \varepsilon \frac{\lambda_F l_{ee}^2(T)}{L^3}
\frac{\varepsilon_F\phi}{T}\; .
\end{equation}

According to Eqs.\ (11) and (14), the signal decreases with
increasing $L$ according to a power-law, but not exponentially.
This again is essentially a two-dimensional effect (cf.\
\cite{Gprb,Glt1}) and should be well-pronounced in high-mobility
samples with sufficiently large $L$.

Thus, it is possible to create conditions in a two-dimensional
electron-gas under which the electron-beam signal is determined
rather by a higher-energy quasi-ballistic group of electrons which
experience small-angle scattering than by purely ballistic
electrons $\varepsilon \le \varepsilon_0$.

\section{Non-linear effects and heating}\label{sec_heat}

Due to the heating of the electron gas between injector and
detector point-contacts for 'high' excess energies the detector
signal consists not only of quasiballistic beam electrons but also
of an isotropic signal resulting in a thermovoltage across the
detector. This causes the growth of $V_d$ for injection energies
$V_i > 5$ mV in our experiments [Fig.\ \ref{exp} a) and b)]. The
contribution of the thermopower to the detector signal is given by

\begin{equation}
\Delta V_d= S(T)\Delta T, \hspace{6mm} \Delta T = T-T_0.
\end{equation}

Here, $T$ is the electron gas temperature between injector and
detector and $T_0$ is the gas temperature beyond the detector
(which is close to the lattice temperature), $S(T)$ is the Seebeck
coefficient (thermopower) of the detector (heating of the 2DEG
between injector and detector by the injected electron beam was
already discussed by us in Ref.\ \cite{Msst}). As discussed above
(Sec.\ \ref{sec_exp}), on increasing $V_i$, the increase of $T$
leads to an increase of the thermovoltage on the one hand, and to
the decrease of the mean-free-path of quasiballistic electrons on
the other hand, and therefore to the appearance of minimum in the
dependence $V_d(V_i)$.

For $\varepsilon_0 >\varepsilon^{*}$ (the limit where specific 2D
effects can be neglected), only ballistic electrons contribute to
the beam signal. We then have as a rough estimate for the
temperature dependence of the signal:

\begin{equation}
V_d(T) \sim {\kappa e^{-\frac{L}{l_{ee}(T)}}} +S(T)\Delta T, \;
\kappa^{-1}\sim L\lambda_F^{-1}\phi \left(
V_i^{-1}+\varepsilon_0^{-1}\right).
\end{equation}

An analysis of this expression shows that a minimum in $V_d(T)$ is
always present when $V_i > \varepsilon_0 $. This statement holds
as well in the multi-collision regime $\varepsilon_0 < T <
\varepsilon^{*}$ [see Eq.\ (14)]. It is evident that with
increasing $V_i$ the temperature of the 2DEG between injector and
detector point-contact increases. However, this does not
necessarily imply that the $V_d(V_i)$ dependence replicates
$V_d(T)$ of Eq.\ (16) qualitatively, because $V_i$ enters
explicitly in Eq.\ (16) and not only through $T(V_i)$. One can
only state definitely that the beam signal should decrease sooner
or later on increasing $V_i$.

Generally speaking, the theoretical determination of the
dependence $T(V_i)$ requires solving a complex non-linear problem
on the beam's self-action. However, the essential dependence of
the mean-free-path on excess energy allows considerable
simplifications for sufficiently high $V_i$, i.e., the separation
of the injected particles in (i) "heating" (high-energy electrons
which do not reach the detector) and (ii) quasiballistic electrons
(which contribute mainly to the beam signal, but not to heating).
Such a separation is undoubtedly possible at $V_i >\varepsilon_0
>\varepsilon^{*}$. (If $V_i \le \varepsilon_0$ we can neglect
heating.)

Thus, under certain conditions one can use the following
quasi-linear approach: First, we find the electron gas heating
$\Delta T(V_i)$ due to the high-energy part of the beam, and then,
using the electron temperature $T$ thus obtained, we determine the
signal of the quasiballistic part. In particular, in the
relaxation-time approximation we have:

\begin{equation}
f=e^{-\frac{L}{v_x \tau_{ee}(\varepsilon,T)}}
f_0 \left( y-L\frac{v_y}{v_x},{\bf p} \right),
\end{equation}

where $T=T_0 + \Delta T(V_i)$. In fact, this means that the
separation reduces the nonlinear problem to two linear equations.

Finally, we want to consider the case when $\varepsilon_0 < V_i
<\varepsilon^{*}$ where, due to the specific two-dimensional
effects, the injected particles slowly relax their directionality,
but rapidly lose their excess energy \cite{Gprl}. In this limit,
it is not possible to separate heating particles from
quasiballistic ones. The beam signal is proportional to
$l_{ee}^2(T)T^{-1}$ [see Eq.\ (14)] and, hence, it is sensitive to
heating. This essentially nonlinear situation could be realized
experimentally for high quality samples with a large distance $L$
between injector and detector.

\begin{figure}[t]
\begin{center}
\resizebox{7cm}{5cm}{\includegraphics{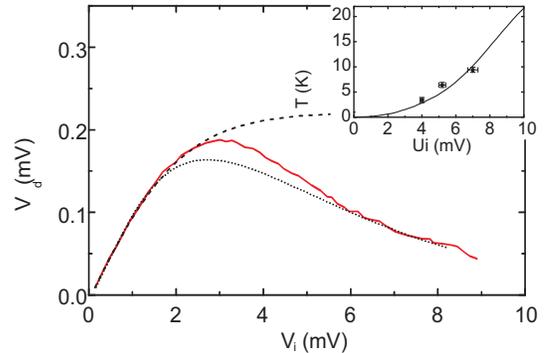}}
\begin{minipage}{8cm}
\caption{\label{theoexp} The comparison of theory and
 experiment for $T_0 = $ 1.6 K: modified one-collision
approximation (dashed line), the experiment (solid line), and the
relaxation-time approximation with taking into account the heating
(dotted line). The inset displays the dependence of the electron
temperature on injector voltage, $\Delta T (V_{i})$, obtained from
heating temperature measurements (solid line) and from the
experimentally determined temperature dependence of the detector
signal (Fig.\ \protect{\ref{exp}}) (squares).}
\end{minipage}
\end{center}
\end{figure}

\section{Discussion of the experiment}\label{sec_discuss}

In this section we want to compare the experimental results, Sec.\
\ref{sec_exp}, with our theoretical results. A series of
measurements for different sample temperatures, $T_0=$ 1.6, 2.2,
3.4, 5.0, 8.0, 11, 15, and 17 K were available for analysis
(partly shown in Fig.\ \ref{exp}). For temperatures $T_0 < 8$ K,
the modified one-collision approximation [Eqs.\ (7), (8) and (9)]
allows a proper description of the experiment for $V_i < 3$ mV,
see Fig.\ 4 (note that $l_{ee} \approx L = 3.4$ $\mu$m for $V_i =
\varepsilon_0 \simeq 2$ mV). For higher values of $V_i$, where
heating is essential, one can use the relaxation-time
approximation [Eq.\ (17)], taking into account the $T(V_i)$
dependence].

To compare theory with experiment we extract the heating $\Delta
T(V_i)$ in two different ways: First, by measuring the heating
caused by the electron beam as a function of injection energy,
using the thermovoltage across the detector quantum point-contact
as a thermometer \cite{thermopower} and second, by analysing the
set of experimental data for the anisotropic part of the signal
(Fig.\ \ref{exp}), see below.

The result of an electron-temperature measurement determined via
thermopower for a lattice temperature $T_0 = 1.6$ K is displayed
in the inset of Fig.\ \ref{theoexp}. Here, the detector
point-contact conductance was adjusted to yield a maximum
thermopower ($S \approx 20$ $\mu$V/K) \cite{thermopower}, where
the conductance of the injector point-contact was fixed at one
mode so that its thermopower is negligible compared with the
detector. The measurements were done at small magnetic fields ($B
= 50$ mT) to prevent beam electrons from reaching the detector
point-contact directly.

Alternatively, the decrease of the detector signal due to the
quasiballistic part of the electron beam allows for an estimate of
the 2DEG beam heating. We assume that the part of these curves at
values $V_i$ larger than the injector voltage at the maximum in
$V_d$ (which we now denote as $V_i^{\rm max}$) describes the
signal from the full narrowly-directed fraction of electrons as a
function of 2DEG temperature, i.e. $V_d(T_0+\Delta T(V_i))$. Thus,
the curves in Fig.\ \ref{exp} are members of a one-parameter
family which differ only by the value of the temperature $T_0$.
For the validity of this statement, it is important that heating
can be neglected at the local maximum of $V_d(V_i)$ for each
curve, $\Delta T(V_{i}^{\rm max}) \approx 0$. From Fig.\ \ref{exp}
b) it is evident that at $V_{i}^{\rm max}$ the thermovoltage is
indeed negligible. Let us now consider any two curves $T_{01}$ and
$T_{02}$ of this family and let $T_{01} < T_{02}$. Then curve
$T_{01}$ has always larger values $V_d$ for a given $V_i$ than
curve $T_{02}$, and curve $T_{01}$ decreases to a signal $V_d$,
equal in size to the maximum of curve $T_{02}$ at a given, larger,
value of $V_i$. Now, it is evident that $T_{01}+\Delta
T(V_i)=T_{02}$. In this manner, we are able to reconstruct the
function $\Delta T(V_i)$. Let us emphasize, that it is convenient
to choose the local maximum $V_{i}^{\rm max}$ as a starting point
for recovering $\Delta T(V_i)$, since, in the vicinity of this
point there is no need (i) to correct for the increase of the
signal due to the quasiballistic group of electrons with
increasing $V_i$, that takes place at low $V_i$ in the linear
response regime, and (ii) to take heating effects into account. As
an example, let us consider the curves 1 and 2 of Fig.\ \ref{exp},
i.e., $T_{01} = 1.6$ K  and $T_{02} = 8$ K. The maximum value of
curve 2 is approximately 0.135 mV. The same value of $V_d$ for
curve 1 is reached in decreasing part of the curve at $V_i \sim
5.2$ mV. Thus, we obtain a heating temperature of $\Delta T \simeq
T_{02}-T_{01}=6.4$ K for $V_i \sim 5.2$ mV. The results obtained
for the experiments at the lowest sample temperature $T_{0} = 1.6$
K are shown in the inset of Fig.\ \ref{theoexp} (squares). It can
be seen that the extracted heating temperatures agree well with
2DEG temperature measurements for applied magnetic fields. We
therefore can use this $T(V_i)$-dependence for further
considerations. Note that the heating temperature depends not only
on $V_i$ but also on the initial sample temperature $T_0$: $\Delta
T = \Delta T(V_i,T_0)$. For higher $T_0$, the 2DEG heating is less
efficient, cf.\ Fig.\ \ref{exp}.

Additionally, the electron temperature can be estimated roughly
from the heat-balance between the energies transfered from the
electron beam into the 2DEG and removed by phonons \cite{Mprl}. We
then have

\begin{equation}
v_F <\varepsilon> \lambda_F \frac{V_i}{\varepsilon_F}n_e \sim
\nu_{ep} s \hbar k_F \frac{T}{\varepsilon_F} n_e \Sigma.
\end{equation}

Here, $<\varepsilon> \sim V_i/2$ is the average energy of the
electron beam, $n_e V_i/\varepsilon_F$ is the number of injected
electrons, $\nu_{ep}$ is the frequency of electron-phonon
collisions, $s$ is the sound velocity, $k_F$ is the Fermi wave
vector, and $\Sigma$ is the area of the heated 2DEG region. Thus,
we obtain for the electron temperature

\begin{equation}
T\sim \frac{\lambda_F l_{ep}}{\Sigma}
\frac{\varepsilon_F}{s\hbar k_F} <\varepsilon>.
\end{equation}

In order to evaluate this expression we assume for the
experimental situation the following values: $s \sim 6 \times
10^5$ cm s$^{-1}$, $k_F = 1.18 \times 10^6$ cm$^{-1}$,
$\varepsilon_F = 9$ mV, and $\Sigma$ is taken to be of the order
of the area between injector and detector, viz.\ 200 $\mu$m$^2$.
The mean-free-path for electron-phonon collisions $l_{ep}$ is
estimated at 100 $\mu$m, yielding $T(V_i = 5.2$ mV) $\simeq 7.3$ K
[Eq.\ (19)]. In spite of this very crude model, we thus find a
remarkable agreement with the electron temperature obtain from the
experimental data ($\Delta T(V_i \simeq 5.2$ mV$) = 6.4$ K).

As mentioned above, the experimental data can be approximated
using the modified one-collision approximation [Eq.\ (7)] for
injection energies $V_i < V_i^{\rm max}$  and the relaxation-time
approximation [Eq.\ (17)] for $V_i$ which is sufficiently large in
comparison with $ V_i^{\rm max}$. At high $T(V_i)$ the scattering
is not small-angular and leads to a more or less isotropic
background, i.e.\ 2D effects are absent. According to this, we
plotted in Fig.\ \ref{theoexp} the modified one-collision
approximation (dashed line), experiment (solid line) for $T_0=1.6$
K and the relaxation-time approximation (RTA) (dotted line), which
takes into account electron-heating effects. For the RTA we have
use the asymptotic expression

\begin{equation}\label{asymp}
  \tau_{ee}^{-1}(\varepsilon, T)= \frac{\varepsilon^2 + 2 \pi^2
T^2}{4 \pi \hbar \varepsilon_F} \ln
\frac{\varepsilon_F}{T+\varepsilon}\; ,
\end{equation}

which is valid for arbitrary ratio of small values of $T/
\varepsilon_F$ and $\varepsilon/\varepsilon_F$ \cite{Glt1}. The
coefficients for the theoretical calculations were chosen in such
a way that coincidence is achieved for small injection energies,
where the linear increase is observed and electron travel
ballistically in any case. Thus, in fact, no additional fitting
parameters are used. For this calculation, it is important that
the detector size $\lambda_F \ll L\phi$. As one can see from Fig.\
\ref{theoexp}, the divergence between the relaxation-time
approximation and the experiment is only due to specific
two-dimensional effects and reaches a maximum in the vicinity of $
V_i = 3$ mV, where on the one hand a number of the scattered
particles is comparable with the number of pure ballistic, and on
the other hand scattering is still small-angular.

\section{Conclusions}

We have studied the role of different groups of electrons on the
propagation of electron beams in a high-mobility two-dimensional
electron-gas for wide range of excess energies. We have observed a
non-monotonic dependence of the detector signal on the excess
energy of the injected electrons. This result can be explained in
the framework of our model, separating the beam electrons into two
groups, i.e. "quasiballistic" electrons and "heating" electrons
(high-energy part of a beam). We have shown that due to the
reduced dimensionality of the system the quasiballistic fraction
consists not only of purely ballistic electrons but also of a
significant number of electrons which have experienced small-angle
electron-electron scattering events. The small-angle character of
electron-electron scattering is essentially a two-dimensional
effect, predicted earlier by us \cite{Gprb,Glt1,Glt2}, which
manifests itself in the experiments discussed here. In addition,
we have formulated the conditions where 2D effects can be best
observed and thus electron-beam propagation over very long
distances should be possible.

\acknowledgements

This work was supported by the Volkswagen-Stiftung (Grant No. I/72
531), and by the DFG MO 771/1-2.

%\begin{thebibliography}{99}

%%%%%%%%%%%%%%%%%%%%%%%%%%
\end{multicols}
%%%%%%%%%%%%%%%%%%%%%%%%%%%


\begin{references}
\bibitem{thermo} Note that while the thermopower of a quantum point-contact in
linear response reaches a maximum in-between conductance plateaus,
and is close to zero near the middle of a plateau
\protect{\cite{Mprl}}, one expects through thermal broadening
effects also a sizeable thermovolatge contribution for the
adjustment of our detector point-contact at a large enough
temperature difference
\end{references}
\end{document}